\documentclass{article}
\usepackage[a4paper, margin=2.5cm]{geometry}
\usepackage{amssymb}
\usepackage{amsmath}
\usepackage{graphicx}
\usepackage{subcaption}
\usepackage{svg}
\usepackage{hyperref}

\begin{document}

\title{Hybrid ILM-NILM Smart Plug System}

\author{Dániel István Németh\footnote{Faculty of Information Technology and Bionics, Pázmány Péter Catholic University}, Kálmán Tornai\footnote{Faculty of Information Technology and Bionics, Pázmány Péter Catholic University}}

\date{}

\maketitle

\section*{Keywords}
\label{sec:keywords}
Electrical Load Classification; Smart Plug; Intrusive Load Monitoring

\section*{Abstract}
\label{sec:abstract}

Electrical load classification is generally divided into intrusive and non-intrusive approaches, both having their limitations and advantages. With the non-intrusive approach, controlling appliances is not possible, but the installation cost of a single measurement device is cheap. In comparison, intrusive, smart plug-based solutions offer individual appliance control, but the installation cost is much higher. There have been very few approaches aiming to combine these methods. In this paper we show that extending a smart plug-based solution to multiple loads per plug can reduce control granularity in favor of lowering the system's installation costs. Connecting various loads to a Smart Plug through an extension cord is seldom considered in the literature, even though it is common in households. This scenario is also handled by the hybrid load classification solution presented in this paper.

\section{Introduction}
\label{sec:introduction}

Renewable energy production sources are seen as the clean solution to the world's energy problems, but the energy grid is becoming a clear bottleneck for the transition. The research community proposed many solutions to tackling the challenge of balancing the electrical grid with non-controllable renewable energy sources present. One such solution is Demand-Side Management (DSM) aiming to adjust the electricity demand to meet the grid's production capacity. 

On the household level, Intrusive Load Monitoring (ILM) and Non-Intrusive Load Monitoring (NILM) have been the two main approaches to gathering load composition data and, in the case of ILM, controlling appliances and thus influencing the demand curve. Both have advantages and disadvantages that limit their real-world applicability.

In this paper, we present a smart plug-based hybrid approach that aims to solve the limitations of Intrusive Load Monitoring by adding multi-load detection capabilities. By detecting multiple appliances connected simultaneously to a single outlet, fewer plugs are required, lowering the installation cost significantly. In addition, plugging multiple loads into a smart plug through an extension cord is also handled, solving a real-world problem of Smart Plugs not yet considered in ILM research.

The rest of the paper is organized as follows. Section \ref{sec:motivation} introduces the relevant literature, and Section \ref{sec:extending} introduces our proposed hybrid ILM-NILM solution. The data collection methodology and dataset are presented in Section \ref{sec:dataset}, followed by Section \ref{sec:why} showing why multi-load detection is necessary for Smart Plugs deployed in real-world environments. Section \ref{sec:mcc} presents the classification approach we chose, and the discussion of the results is shown in Section \ref{sec:discussion}. Finally, the conclusions are presented in Section \ref{sec:conclusions}.

\section{Context and motivation}
\label{sec:motivation}

With the electricity generation from solar and wind power expected to grow significantly in the following decades \cite{eiaInternationalEnergy}, maintaining grid stability with non-controllable resources on the production side is becoming more and more challenging. As the world transitions to renewable energy sources, several solutions have been proposed to solve the energy demand timing problem. To support stable grid operations, load composition data is crucial to predict future grid load \cite{BISWAL20243654}.

Electrical Load Classification is generally divided into Non-Intrusive and Intrusive Load Monitoring (NILM and ILM). Both approaches have limitation that make it difficult to apply them in real-world scenarios.

With NILM, electricity consumption is measured at a single point in the household, utilizing a Smart Meter \cite{10116809}. It is an inexpensive way to collect aggregated consumption data, but additional disaggregation steps are required to infer appliance-level consumption data. Several methods have been proposed in the literature for disaggregation, ranging from Hidden Markov Models \cite{9967943} and Mixed Integer Linear Programming \cite{10324503} to Neural Network-based classification \cite{Ouzine2023} and genetic algorithms \cite{en15239073}. In addition to machine learning methods, pattern matching, and source separation-based NILM disaggregation solutions have also been proposed in the literature \cite{9820770}. With NILM, there is no control possibility, so it relies on manual appliance control for any energy savings or control \cite{9166709}. Detection of multi-state loads is another challenge most NILM solutions struggle with. Electrical appliances can be categorized into four types based on their load patterns:
\begin{itemize}
    \item Type I. On/off appliances
    \item Type II. Finite State Machine-like multi-state appliances
    \item Type III. Continuously variable power appliances
    \item Type IV. Permanently on appliances \cite{s121216838}
\end{itemize}
Type I. and IV. loads are easier to identify, while only a few state-of-the-art NILM approaches focus on appliances with variable power usage \cite{9906271, 10324503}. Detecting loads with variable energy usage is a challenge for ILM too \cite{6977348}.

With ILM, each appliance is measured separately using Smart Plugs. It offers accurate, appliance-level consumption data, load classification, and load control capabilities but at much higher installation costs, as a separate Smart Plug needs to be installed for each appliance. Utilities can indirectly influence energy consumption by using different pricing schemes \cite{compreviewofdsm}. Still, the high costs and limitations of Smart Plugs limit their commercial application even though appliance-level power consumption data could be used for activity monitoring \cite{8448304}, occupancy detection and automated appliance control \cite{TEKLER2022109472}.

In addition to low-frequency time-series-based ILM, recent approaches have been using high-frequency measurements, offering faster detection times. Voltage--Current trajectories from instantaneous voltage and current reading can be used to categorize the connected electrical load \cite{7130652}. In \cite{8756885}, Electric Load Signatures are recorded for electrical loads used to classify the load. Manipulating the electrical load's supply to extract characteristic information is another way to achieve fast ILM \cite{8086053, ccis, energiesopenset}, and this paper extends that approach.

\subsection{Hybrid approaches}

Both ILM and NILM have shortcomings that limit their real-world application. Smart Meters (NILM) lack control capabilities, and disaggregation algorithms have their challenges, especially when it comes to appliances with variable power usage. For Smart Plugs, there is a high installation cost as every appliance needs to have its own Smart Outlet. Plug loads can be moved within the household, so each appliance needs to collect enough information to classify the connected load correctly.

There have been only a few novel approaches of combining ILM and NILM in different ways \cite{10.1145/3208903.3212052}. In \cite{7436413}, NILM is extended with controllable outlets and an automatic appliance mapping algorithm, which aims to solve the lack of control capabilities of non-intrusive approaches, but this negates most of the cost savings of NILM compared to ILM. An other approach, Semi-Intrusive Load Monitoring (SILM), is proposed in \cite{7210212}. SILM is essentially NILM, but with several measurement points instead of one. The goal is to assign electrical appliances to a minimal number of groups such that one load meter per group is enough to recover the state of any appliance. However, their method has limitations as follows. The appliances are assumed to be only types I. (on/off), II. (finite states) or IV. (permanently on), loads with continuously variable power consumption (e.g., any device with a built-in battery) are not considered. In \cite{10.1145/3208903.3212052} it is shown that the presence of appliances with continuously variable power consumption characteristics significantly decreases NILM performance, and Völker et al. proposed recording load-level data for Type III. loads to improve NILM performance. This hybrid approach, however, does not consider the fact that Type III. loads are ubiquitous in households (e.g.,  battery-powered devices), and these may be plugged in to any outlet. \textbf{Our approach, presented in the next section, is different than any of the above hybrid approaches and solves multiple critical shortcomings of Smart Plugs.}

\section{Extending Smart Plugs with multi-load detection capabilities}
\label{sec:extending}

This paper presents a different, novel hybrid approach, focusing on two significant limitations of Smart Plugs. Firstly, Smart Plug installation costs are high as each appliance needs to have its own plug. Additionally, the literature does not consider the scenario of connecting an extension cord and, thus, several loads to a single Smart Plug. For Smart Plugs to be applicable in real-world scenarios, both of these limitations need to be overcome. The hybrid approach presented in this paper solves both of these limitations by extending the load detection capabilities of dimmer-based Smart Plugs \cite{8086053, ccis, energiesopenset} to multiple loads connected to a single Smart Outlet. 

Smart Plugs capable of detecting multiple appliances connected in parallel would solve several limitations of ILM-based approaches. If there is no need for each appliance to have its own outlet, the cost of such a system is significantly lower. In addition, smart plugs can be built into wall outlets as plugging in several appliances through an extension cord is no longer an issue. Creating a Smart Plug capable of detecting multiple loads simultaneously solves the misclassification issues arising when multiple loads are connected to an outlet designed to detect a single load only. We demonstrate this in Section \ref{sec:why}. 

However, some constraints have been introduced. If multiple appliances are connected to a single Smart Plug, the loads can only be controlled together. If either of the loads is non-deferrable, the outlet cannot be turned off automatically.

\begin{figure}[t]
\centering
\includegraphics[width=260pt]{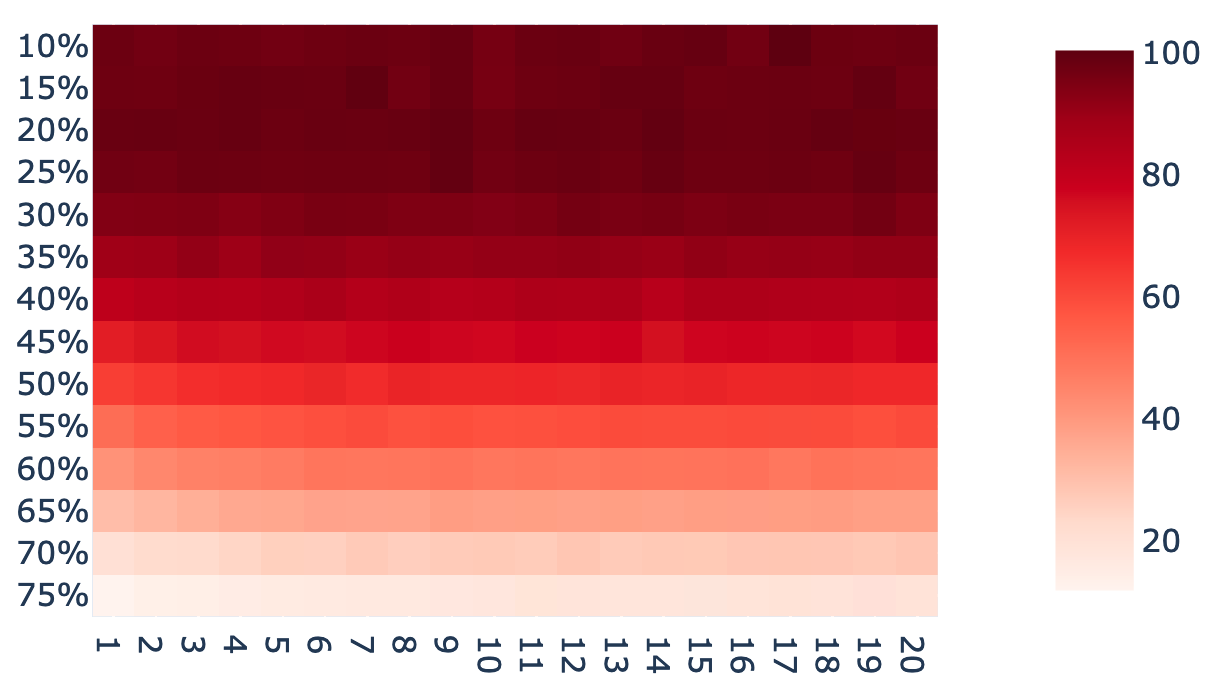}
\caption{Real Power measurement matrix [W] of an incandescent light bulb. Each row represents a different cutoff ratio, while each cell shows the Real Power measurement recorded for a single AC period.}
\label{fig:incandescentexample}
\end{figure}

With most Smart Plugs in the literature measuring appliance power characteristics with low-frequency, we opted to extend the dimmer-based approach published in \cite{ccis, energiesopenset} due to its high-frequency data collection and fast load classification capabilities. The Smart Plug extracts three $14 \times 20$ matrices within 10 seconds from the connected electrical load by cutting the voltage supply at various ratios. The three matrices containing the RMS voltage, RMS current, and real power measurements are populated in a row-major order. First, the load's supply is dimmed with a 10\% cutoff ratio for 20 AC periods. The matrices' first rows represent the measurements recorded for each dimming cycle. Then, after a brief period of uninterrupted power, the cutoff continues with 15\%, measuring data for the 2nd row. The cutoff ratio is increased by 5\% for each row, resulting in 75\% for the last row. The matrices recorded in this way can be used to identify the connected load \cite{ccis} accurately. The real power matrix for an incandescent light bulb is shown in Figure \ref{fig:incandescentexample} as an example.

To the best knowledge of the authors, the hybrid approach of integrating multi-load capabilities into Smart Plugs has not been examined in the literature before.

\section{Data collection}
\label{sec:dataset}

\renewcommand{\arraystretch}{1.5}
\begin{table}[]
\centering
\begin{tabular}{|l|c|l|}
\hline
Appliance description                                                                                   & Label               & Load category \\ \hline
\begin{tabular}[c]{@{}l@{}}5W and 10W USB adapters \\ charging portable electronic devices\end{tabular} & USB                 & Type III.     \\ \hline
\begin{tabular}[c]{@{}l@{}}4A "smart" lead-acid \\ battery charger\end{tabular}                         & batterycharger4A    & Type III.     \\ \hline
\begin{tabular}[c]{@{}l@{}}800mA traditional \\ lead-acid battery charger\end{tabular}                  & batterycharger800mA & Type III.     \\ \hline
A fan                                                                                                   & fan                 & Type II.      \\ \hline
A hairdryer                                                                                             & hairdryer           & Type I.       \\ \hline
LED Light Bulb                                                                                          & ledbulb             & Type I.       \\ \hline
LED-based Spotlight                                                                                     & ledspotlight        & Type I.       \\ \hline
\begin{tabular}[c]{@{}l@{}}Incandescent light bulbs\\ for  lighting and heating\end{tabular}            & INCANDESCENTS       & Type I.       \\ \hline
\begin{tabular}[c]{@{}l@{}}Laptop charger\\ charging a laptop\end{tabular}                              & laptop              & Type III.     \\ \hline
LCD screen                                                                                              & monitor             & Type I.       \\ \hline
Soldering iron                                                                                          & solderingiron       & Type I.       \\ \hline
\end{tabular}
\caption{Electrical appliances measured, their categorization (appliance type) and their labels used for reference in this paper.}
\label{table:loads}
\end{table}

When deciding on the appliances to measure, we considered two important facts. Firstly, only loads moved within the household should be examined. An installed HVAC system or a freezer may have high energy consumption, but load detection is not required as it is not moved between outlets. In contrast, laptops, USB chargers, and other appliances like lamps can be plugged into any outlet, so our dataset should reflect the types of loads commonly moved between outlets. Secondly, as highlighted in \cite{6977348}, traditionally, Type I. and IV. (on/off and permanently on) loads are easier to detect, while Type II. and III. (FSM-like and continuously variable) loads are more difficult to classify correctly. As many plug loads belong to the latter categories, we need to include these types of loads, too, despite the challenges involved.

We have collected measurement data from a total of 14 electrical loads, belonging to 11 distinct appliance types shown in Table \ref{table:loads}. As can be seen in the table, we opted to measure several loads with continuously variable power consumption characteristics. The power consumption of these Type III. loads may be influenced by human activity (e.g., laptop power usage) or by their built-in energy storage devices (e.g., phone's battery, lead acid battery). Table \ref{table:loads} also shows the load labels that are used in further figures to refer to these devices.

\begin{figure}[t]
  \centering
   \includegraphics[width=300pt]{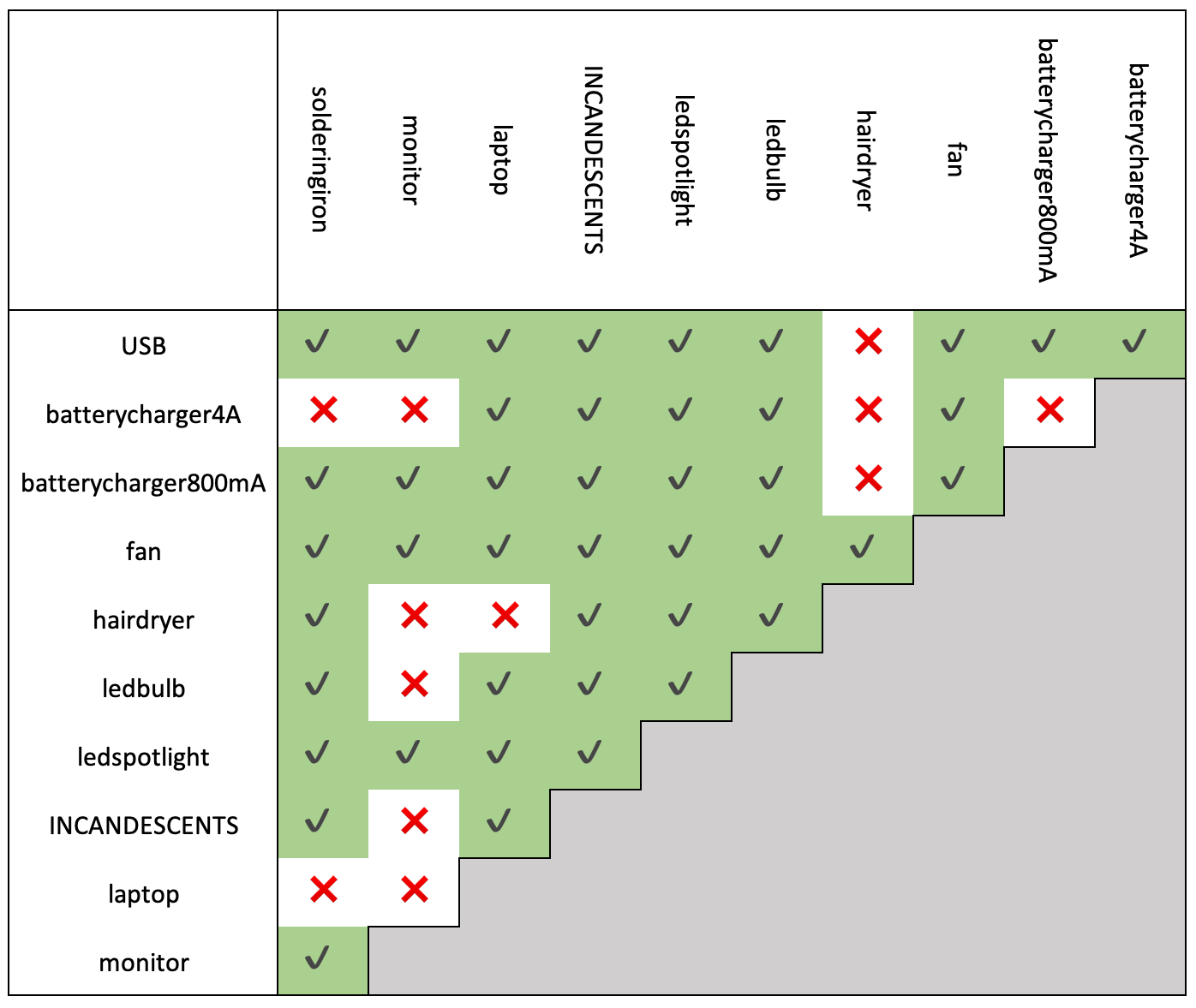}
  \caption{All possible two-load category combinations. $\checkmark$ marks the combinations for which at least 100 measurements were recorded.}
  \label{fig:comb2}
\end{figure}

To evaluate our method, it is not enough to gather measurements from each single load. To assemble a dataset, we would need to measure every possible combination of loads. That, however, would result in lots of measurements. In fact, to gather data only for n-load combinations of k distinct loads, ${n \choose k}$ different combinations would need to be measured. For our 11 appliances, it would involve measuring 55 two-load combinations and 165 three-load combinations. Out of these, we opted to test our method by gathering measurements from 43 of the 55 two-load combinations and from an additional 2 three-load combinations. Together with the single-load measurements, we have assembled a dataset consisting of the following:
\begin{itemize}
    \item 250 measurements for each of the 14 appliances
    \item 100 measurements for the 43 two-load combinations shown in Figure \ref{fig:comb2}
    \item 100 measurements for the \textit{(fan + ledbulb + INCANDESCENTS)} and \textit{(fan + ledspotlight + solderingiron)} three-load combinations
\end{itemize}
With the above measurements, we have assembled a dataset with a total of 8000 measurements.

\section{Motivating example}
\label{sec:why}

\begin{figure}[]
  \centering
   {\includegraphics[width=\columnwidth]{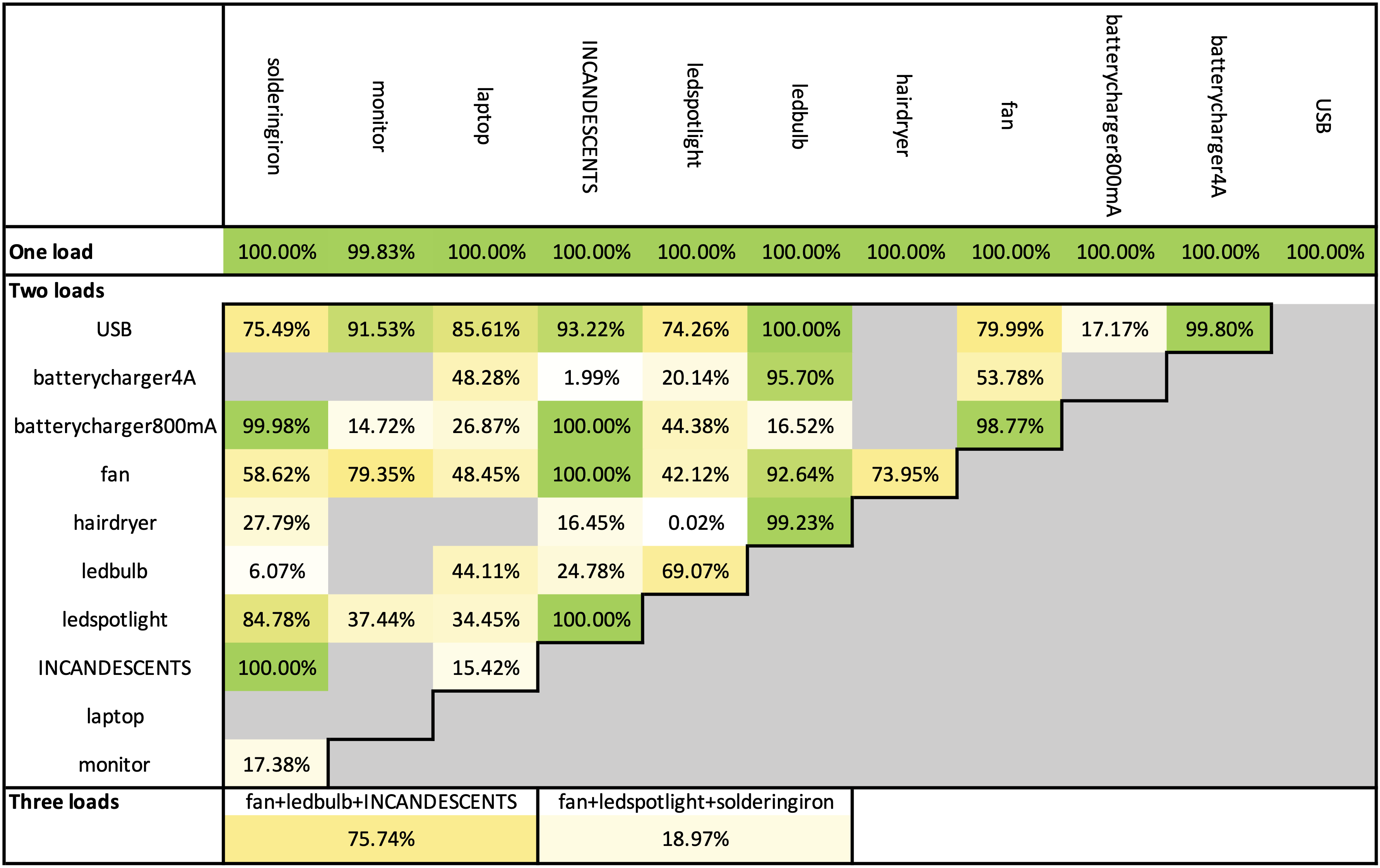}}
  \caption{Traditional classification accuracy rates (average of 100 runs) selecting the top prediction of the CNN. For a multi-load sample, the classification was considered correct if the top output returned was one of the several appliances connected to the plug. The training data did not contain any multi-load samples. The training data consisted of 220 single-load measurements from each load category. The first row contains the accuracy rates for the 30 test samples per load class, the table in the middle contains the accuracy rates for the two-load combinations, and the bottom two accuracy rates belong to the three-load combinations. }
  \label{fig:single2multi}
\end{figure}

\begin{figure}[]
  \centering
   {\includegraphics[width=\columnwidth]{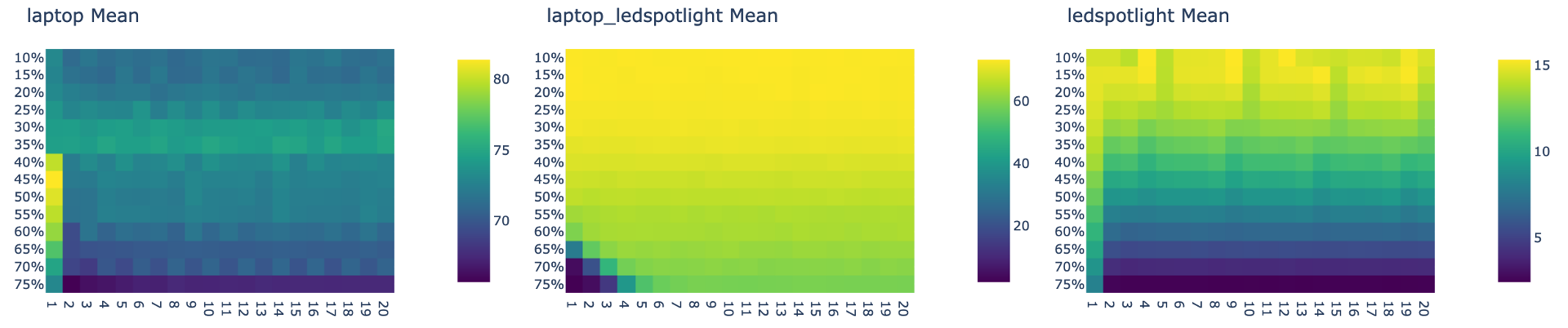}}
  \caption{The mean of all measured Real Power measurement matrices for the laptop and ledspotlight loads (left and right figures) and the mean Real Power measurement matrix of the combined laptop + ledspotlight measurements (middle figure).}
  \label{fig:laptopledspot}
\end{figure}

With the dataset collected, we can examine what would happen if several electrical loads were connected to a Smart Plug designed for detecting only a single appliance. For that, a CNN-based classifier was trained based on 220 randomly selected single-load measurements from the appliance class. Then, the measurements of the 30 remaining single-load samples and all the multi-load measurements (100 per load combination) were classified using the trained model. Repeating these steps a total of 100 times, the results are shown in Figure \ref{fig:single2multi}. As we can see, for single-load measurements, an average of 99.98\% accuracy rate was achieved. In comparison, for the multi-load samples, the average accuracy rate is 57.89\% with accuracy rates as low as 0.02\% for certain combinations. The results are bad, especially since for multi-load measurements, a sample was considered correctly classified when the predicted class was one of the loads in the load combination. 

As can be seen in Figure \ref{fig:single2multi}, there were some combinations where at least one of the load was detected correctly, but for a significant portion of the load combination measurements, the model predicted a completely different load than either of the loads connected to the plug. These results highlight how the performance of traditional ILM-based load classification methods may degrade when connecting more than one load to a Smart Plug. 

The results demonstrate that relying on only single-load detection in a real-world scenario would result in poor performance. The foremost reason for the poor performance is that when measuring multiple loads simultaneously, the loads are connected (even during dimming) and act as one circuit. As such, the measurement is not simply a sum of two separate measurements but a measurement of this new circuit formed by connecting the two loads in parallel. Both loads can have capacitive and/or inductive characteristics. It can be seen in Figure \ref{fig:laptopledspot} where the means of all Real Power measurement matrices are shown for the single-load measurements for the \textit{laptop} and \textit{ledspotlight} loads and the multi-load measurements where the two loads are measured simultaneously. It is worth noting that the laptop's charge cycle and usage may have resulted in different mean power consumption values, but this does not significantly affect the relative structure of these mean matrices. 

\section{Multi-class classification}
\label{sec:mcc}

For a traditional appliance detection algorithm, only one output, the label of the correct load, is expected. This is usually achieved by a classification algorithm, where the output is the prediction of the connected load. This approach needs to be extended to enable the detection of multiple loads simultaneously.

Multi-label predictions are common in NILM \cite{7498597}. The simplest method is to create a label power set and train a single-class classifier with this power set as the label for the output. With this method, however, the number of classes to distinguish becomes very large, and there may be combinations of loads for which no training data is available. There have been other proposed methods such as RAkEL \cite{10.1007/978-3-540-74958-5_38}, Random Forests \cite{pr7060337}, ML-kNN \cite{ZHANG20072038} and Neural Networks \cite{1683770}.

Since a CNN-based Neural Network performed the best on the single-load classification with our dimmer-based data collection previously, we aimed to find a way to extend the Neural Network-based approach with multi-class classification capabilities.

In most applications of CNN, the Softmax activation function is used after the last layer of the network, resulting in a probability distribution, of which the most likely class is chosen. The target output the network would be trained to learn looks something like this: $[0,0,0,0,\mathbf{1},0,0,0,0,0,0]$. This results in the smallest error if the 5th neuron's output (before the Softmax activation) is significantly larger than any other neuron's in the last layer. We can modify this target to expect multiple neurons to have strong activation: $[\mathbf{.5},0,0,0,0,\mathbf{.5},0,0,0,0,0]$. For this target, the output error is the smallest when the 1st and 6th neuron's output is equally high (before the Softmax activation) compared to any other neuron's output in the final layer. This way, both neurons learn to fire when the input is a multi-load measurement. This ensures that if the network learned the data correctly, then for an N-load multi-load input matrix, the top N neurons with the most significant output should be the N loads present in the multi-load input. 

However, the above method leaves an important question: how can we determine N? A threshold learning algorithm can be used for that. To minimize computational complexity, we wanted the same Neural Network to learn to predict the number of loads present in the measurement by adding three additional neurons to the final layer. The first of these neurons should fire if only one load is present, the second if there are two loads, and the third if there are three loads. This way, the network predicts N (the number of loads), and we can select the top N neurons from the first part of the output layer. So, for example, the target output for some training samples may look like:
\begin{itemize}
    \item One-load measurement: $[0,0,0,0,\mathbf{1},0,0,0,0,0,0\:|\:\mathbf{1},0,0]$ 
    \item Two-load measurement: $[\mathbf{.5},0,0,0,0,\mathbf{.5},0,0,0,0,0\:|\:0,\mathbf{1},0]$
    \item Three-load measurement: $[\mathbf{.33},0,0,\mathbf{.33},0,\mathbf{.33},0,0,0,0,0\:|\:0,0,\mathbf{1}]$
\end{itemize}
A custom Softmax function was also needed to ensure the training worked correctly: The Softmax activation was calculated separately for the output neurons' first part (load categories) and the second part (number of loads). 

\subsection{Multi-class classification performance measures}

There are several different error measures for multi-class classification methods introduced in the literature \cite{1683770}. Our network predicts two separate variables: 1) the probability distribution of electrical loads and 2) the top N loads to consider in the distribution. So it makes sense to evaluate the performance of these two methods separately as well. For this, we have selected the accuracy measure. An output is considered correct for the distribution if the top N loads are the N loads present in the training sample. For this part, we assumed to know the true N. For predicting the number of loads, we simply compared if the correct neuron corresponding to the true N had the most significant output value. 

To measure the performance of the two parts of the prediction combined, we introduce the \textbf{strict accuracy} measure: a sample is considered correctly classified if both the distribution and prediction of the number of loads are correct, i.e., if the output of the multi-class classification solution matches precisely the expected output. If either the top N loads in the distribution or predicting the number of loads don't match with the true values, the sample is considered incorrectly classified. We call this strict accuracy since it does not reward scenarios when part of the prediction is correct.

\subsection{Evaluation process}
\label{sec:evalproc}

To measure the performance of our hybrid approach, we applied the following process:
\begin{enumerate}
    \item From the dataset presented in Section \ref{sec:dataset}, select 70 measurements from each 1-load, 2-load and 3-load combination randomly to form the training set, and select 30 measurements to form the test set.
    \item Train a Convolutional Neural Network consisting of two convolutional and two hidden fully connected layers. The network utilized softmax-provided normalization between the convolutional and fully connected layers, while all other activation functions were leaky ReLU. The training data of the CNN is the Real Power and Apparent Power measurement matrix for a measurement. The apparent power is calculated by multiplying the RMS voltage and current values for each cell of the measurement matrices. The output of the networks consists of the load category distribution prediction and the load count prediction.
    \item Evaluate the trained model on the test set, calculating the accuracy of the distribution and the strict accuracy for each load combination category.
    \item Repeat the previous steps a total of 100 times.
\end{enumerate}

\section{Discussion of the results}
\label{sec:discussion}

\begin{figure}[]
  \centering
   {\includegraphics[width=300pt]{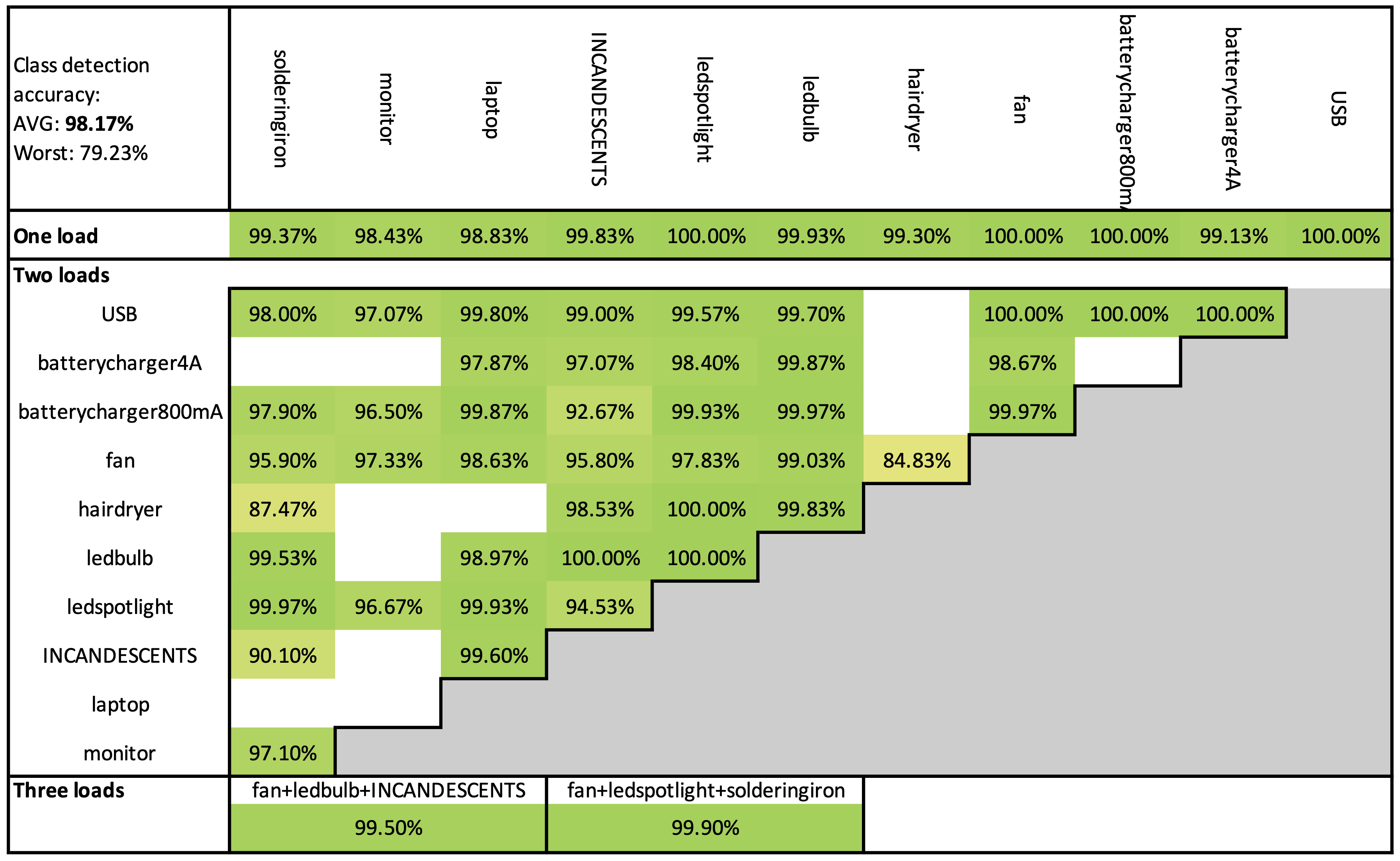}}
  \caption{Summary of class detection accuracy results using 70 training samples from each load combination category and 30 test samples. The accuracy rates are averages of 100 runs. The empty white cells show the combinations for which no measurements were recorded.}
  \label{fig:softresult01}
\end{figure}
\begin{figure}[]
  \centering
   {\includegraphics[width=300pt]{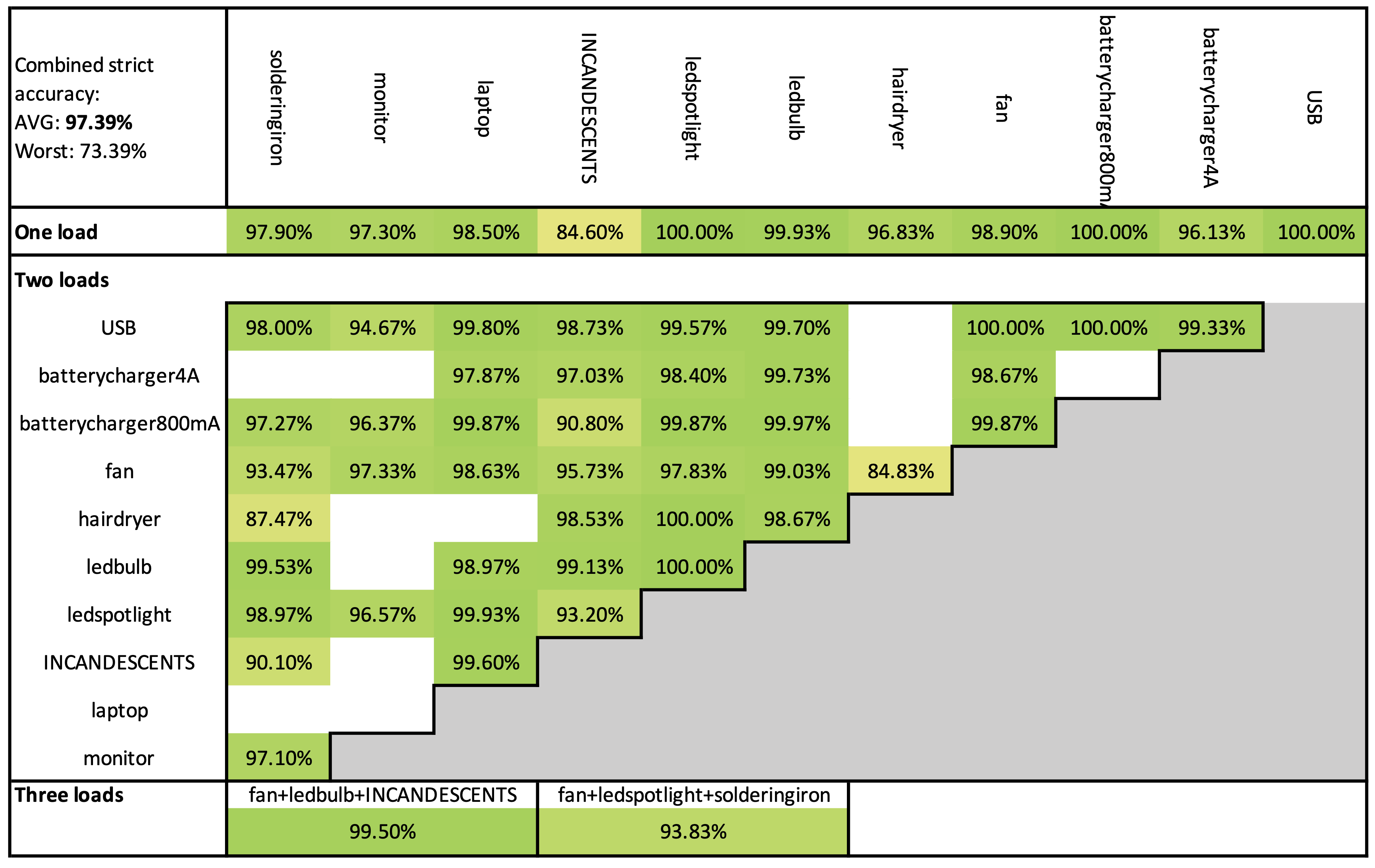}}
  \caption{Summary of strict accuracy results using 70 training samples from each load combination category and 30 test samples. The accuracy rates are averages of 100 runs. The empty white cells show the combinations for which no measurements were recorded.}
  \label{fig:strictresult01}
\end{figure}

Evaluating our hybrid ILM-NILM approach with the process introduced in Section \ref{sec:evalproc}, the results are displayed in Figures \ref{fig:softresult01} and \ref{fig:strictresult01}. Figure \ref{fig:softresult01} shows the accuracy of the first output of the model, the distribution of most likely loads. As can be seen, if the number of loads to expect is assumed, the distribution prediction is correct in most cases, achieving an average accuracy rate of 98.17\%. When the threshold part of model's output is also considered (strict accuracy metric), the accuracy rate is only slightly worse, 97.39\%, as can be seen in Figure \ref{fig:strictresult01}. We can see that accuracy rates are generally lower for load combinations involving higher power loads (e.g. \textit{INCANDESCENTDS}, \textit{hairdryer}) than for combinations of loads with similar power consumption.

\begin{figure}[]
  \centering
   {\includegraphics[width=300pt]{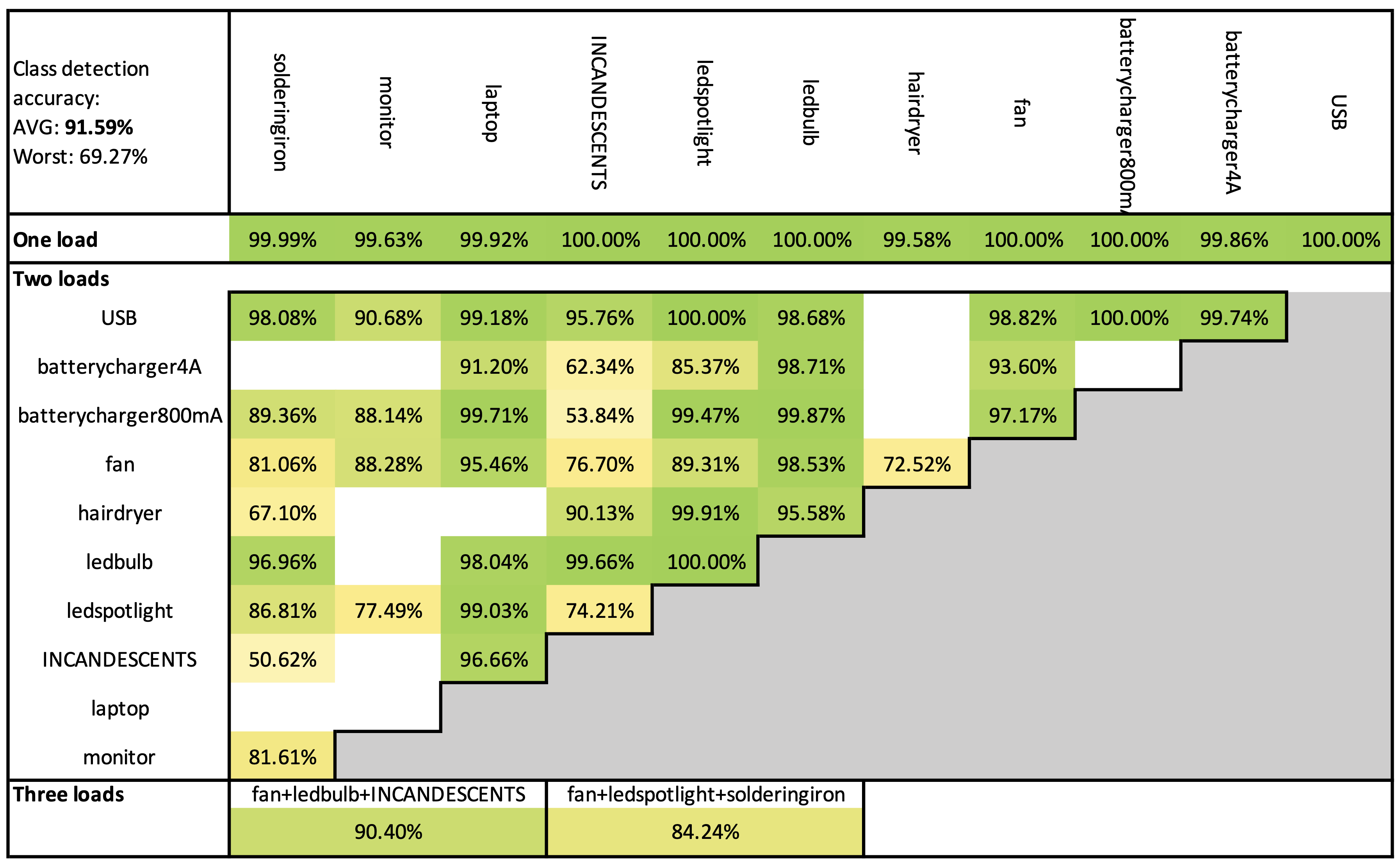}}
  \caption{Summary of class detection accuracy results using 10 training samples from each load combination category and 160 from the single-load categories. The accuracy rates are averages of 100 runs. The empty white cells show the combinations for which no measurements were recorded.}
  \label{fig:mult10_soft}
\end{figure}
\begin{figure}[]
  \centering
   {\includegraphics[width=300pt]{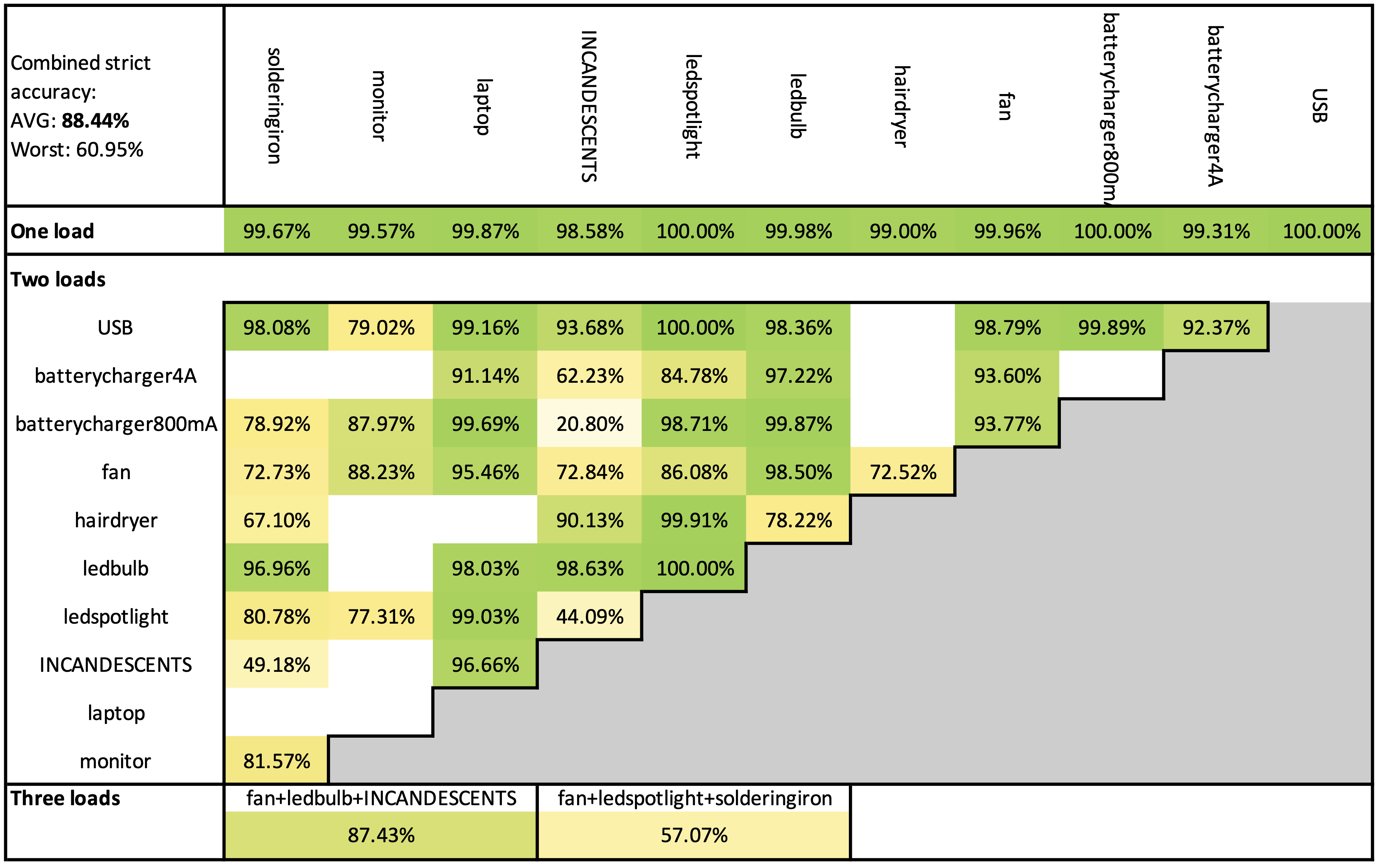}}
  \caption{Summary of strict accuracy results using 10 training samples from each load combination category and 160 from the single-load categories. The accuracy rates are averages of 100 runs. The empty white cells show the combinations for which no measurements were recorded.}
  \label{fig:mult10_strict}
\end{figure}

\begin{table}[]
\centering
\begin{tabular}{|c|cc|cc|}
\hline
& \multicolumn{2}{c|}{Class Detection Accuracy}   & \multicolumn{2}{c|}{Strict Accuracy}   \\ \hline
\begin{tabular}[c]{@{}c@{}}multi-load \\ train samples\end{tabular} & \multicolumn{1}{c|}{AVG}      & Worst    & \multicolumn{1}{c|}{AVG}     & Worst   \\ \hline
70                                                                  & \multicolumn{1}{c|}{98.17\%} & 79.23\%   & \multicolumn{1}{c|}{97.39\%} & 73.39\% \\ \hline
10                                                                  & \multicolumn{1}{c|}{91.59\%} & 69.27\%   & \multicolumn{1}{c|}{88.44\%} & 60.95\% \\ \hline
\end{tabular}
\caption{Class detection and strict accuracy rates (average and worst over 100 runs) when using training datasets of different size.}
\label{table:7010}
\end{table}

With our hybrid approach shown to be able to detect multiple loads accurately, we need to investigate the feasibility of our solution. With our model being trained on 70 samples per combination, a total of $(11+43+2)*70=3920$ samples were used to train the model. With each measurement taking a little under 10 seconds to record, almost 11 hours of continuous measurement would be required to collect a training dataset. To reduce this time, we have two options:
\begin{itemize}
    \item Collect fewer measurements from each combination
    \item Omit the collection of some combinations
\end{itemize}

If reasonable accuracy rates can be achieved with less than 70 measurements per load combination, then it would allow the method to scale better when new loads are added. To test if using less than 70 measurements would be feasible, we used the same method as earlier to train a CNN classifier, but with less training data provided. From each single-load category, 160 measurements were randomly selected for the training set, while the other 90 were allocated to the test set. For the multi-load combinations, for each combination, 10 samples were randomly chosen to be in the training set, while the remaining 90 were assigned to the test set. Similarly to the previous method, this process, along with the model training and test set evaluation, was run 100 times. The classification part's accuracy rates and the strict accuracy rates are shown in Figures \ref{fig:mult10_soft} and \ref{fig:mult10_strict}. We can see cases where strict accuracy rates are low, but the average accuracy rate is still high, 88.44\%. Table \ref{table:7010} shows a comparison of these results and the earlier ones achieved with more multi-load training data available. We can see that although there was a 7-fold reduction in the number of multi-load training samples used, the average accuracy rates are still high. Additionally, by only requiring 10 measurements for each multi-load combination, \textbf{the total measurement time for multi-load samples is about 75 minutes instead of 9 hours.}

\begin{figure}[]
  \centering
   {\includegraphics[width=\columnwidth]{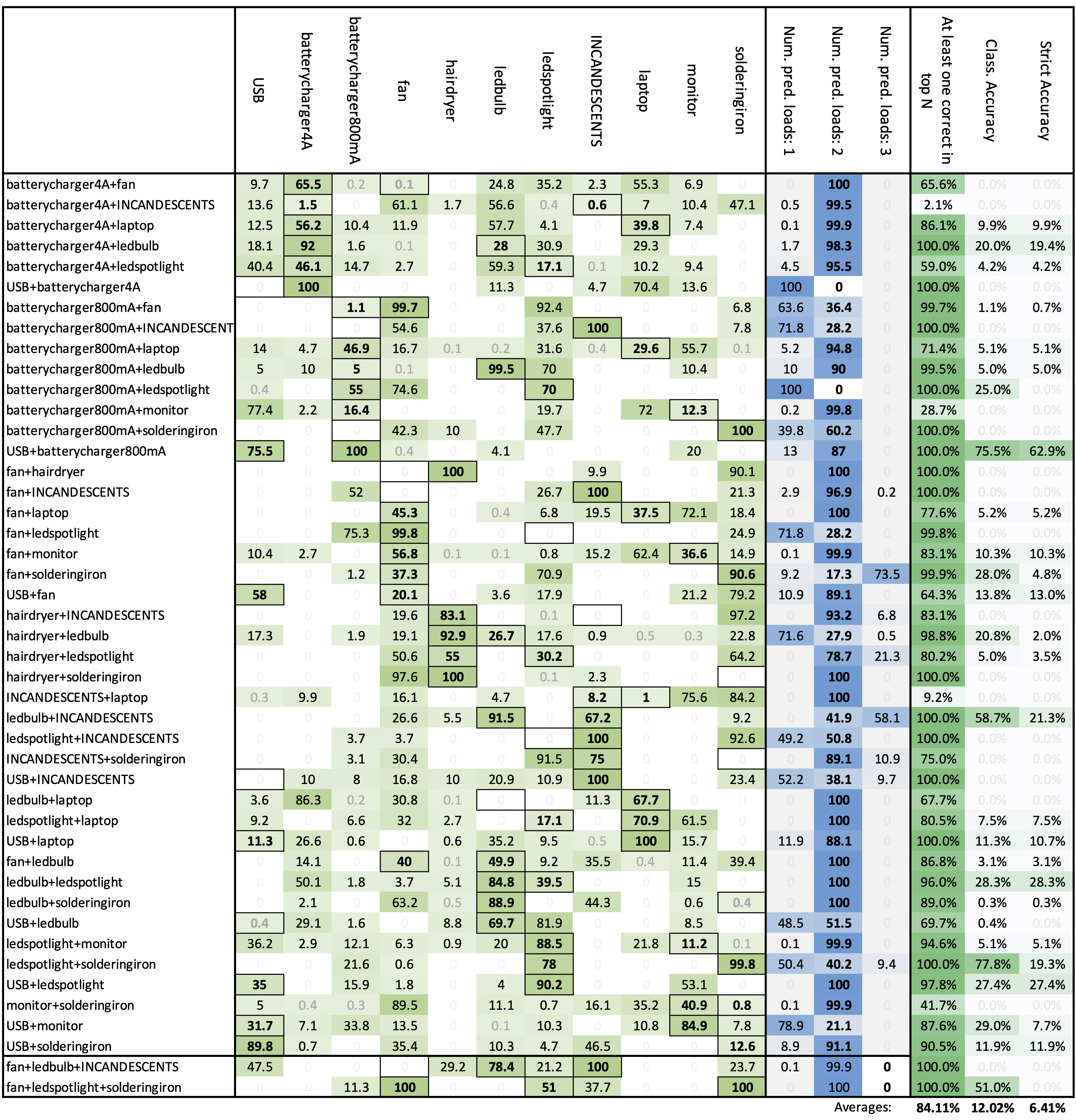}}
  \caption{Multi-load classification results. Each row represents the sample predictions belonging to the omitted multi-load combination. The training set for each row consisted of 70 samples per single and multi-load combinations (except the omitted one). The top N predictions (averages of 10 runs) are displayed in the left green part, the class number predictions in the blue part, and the average accuracy rates for the samples belonging to the omitted class are displayed in the right green part.}
  \label{fig:multi2multi}
\end{figure}

The other approach is to see if our model can still accurately predict some load combinations for which no training data was provided. Still, the loads present in the input have been in the training data both individually and in combination with other loads (e.g., we have the \textit{USB+fan} combination that we omit from the training data; both the \textit{USB} and the \textit{fan} are present in the training data in several other combinations and as single-load measurements as well). By examining how the model classifies an unseen but familiar (i.e., has seen components of it) combination, we understand the model's limitations and capabilities. Figure \ref{fig:multi2multi} displays the average results of this method run ten times for each combination. We can see that in 88.11\% of the cases, at least one load was correct in the top N predictions of the model. In 12.02\% of the cases, all N category predictions of the model were correct. In 6.41\% of the cases, in addition to the N category predictions being accurate, the load count (number of loads) prediction was also correct. These last two percentages are low numbers, which warrant an explanation. Some of the errors certainly come from the network's inability to model the interaction of two loads and their resulting measurement matrix correctly. Another significant portion of the error may come from cases when the model assumes that only one load is present. If we look at the rows in Figure \ref{fig:multi2multi} where the first blue column's (\textit{Num. pred. loads: 1}) values are high, most of the combinations consist of two loads that have significantly different power consumption levels (e.g., \textit{batterycharger800mA+fan}, \textit{batterycharger800mA+INCANDESCENTS}, \textit{USB+INCANDESCENTS}, \textit{USB+monitor}, \textit{USB+batterycharger4A}, \textit{hairdryer+ledbulb}). In these cases, it is understandable that the model predicts only one load and picks the one with the higher power consumption. It is an important observation to consider when collecting multi-load measurement data.

\section{Conclusions}
\label{sec:conclusions}

In this paper we have presented a Smart Plug capable of detecting multiple loads simultaneously. A large dataset was collected to enable the evaluation of our novel method. We have shown that the Neural Network we designed is capable of accurately detecting at least up to three different loads connected to the same Smart Plug simultaneously. With our strict accuracy measure, 97.39\% average accuracy rate was achieved. 

We have also explored the limitations of our solution by examining how the model responds to unseen data related to loads that it has encountered as part of the training data as either part of a different combination or as an individual load. We have also shown that only 10 samples per multi-load combination are enough to reach an average strict accuracy rate of 88.44\%. 

To summarize, we have shown that multi-load detection and, thus, a Smart Plug with hybrid ILM-NILM detection capabilities is possible. With such a solution, we are one step closer to creating a Smart Plug capable of working in a real-world environment. Real-world applicability is rarely considered in the literature, and to the best of our knowledge, we are the first to successfully extend ILM detection methods with load disaggregation, creating a hybrid ILM-NILM method relying only on Smart Plugs.

\subsection{Future work}

With the vast number of potential multi-load combinations, we aim to assemble a larger dataset containing measurements from several additional multi-load combinations. Additionally, in this paper, we did not distinguish between load combinations based on how common they would be in a household. For example, plugging in a hairdryer and fan into the same plug may be less likely than a USB charger and laptop using the same outlet. With a study of such load co-occurrences, the number of likely combinations could be determined, which would be a key asset in assembling a more comprehensive dataset.

We also aim to explore disaggregation methods to determine the power usage of the individual loads from the aggregated measurements, as the current system only detects the loads present but does not estimate the individual power consumption of the loads. If the combination contains more than one appliance with variable power consumption, attributing the power consumption to individual loads becomes challenging.

\section{Funding}
This research was supported by the National Research, Development and Innovation Office (NKFIH) through the grant no. TKP2021-NVA-26.

This work has been supported by the Fund FK 137 608 of the Hungarian National Research, Development and Innovation Office.
\bibliographystyle{ieeetr}
\bibliography{sources}
\end{document}